\documentclass[11pt]{article}

\usepackage{graphicx}
\usepackage{calc}
\usepackage{rotating}
\usepackage[english]{babel}
\usepackage{graphicx}
\usepackage{subfig}
\usepackage{float}
\usepackage{amsmath}
\usepackage{amssymb}
\usepackage{amsthm}
\usepackage{latexsym}
\usepackage{dcolumn}
\usepackage{geometry}
\usepackage{hyperref}
\usepackage{slashed}
\usepackage{multirow}
\usepackage{hhline}
\usepackage{mciteplus}

\usepackage{cite}

\def\gtwid{\mathrel{\raise.3ex\hbox{$>$\kern-.75em\lower1ex\hbox{$\sim
$}}}}
\def\vio{\mathrel{\hbox{$E$\kern-.60em\hbox{$/
$}}}}
\newcommand{\ba}{\begin{eqnarray}}
\newcommand{\ea}{\end{eqnarray}}

\newcommand{\be}{\begin{equation}}
\newcommand{\ee}{\end{equation}}

\begin{document}

\begin{center}
{\Large \bf {TeV-scale leptoquark searches at the LHC \\[0.25cm]
and their E$_6$SSM Interpretation} \\
\vspace*{0.8cm}
{\large Murad Ali$^a$, Shaaban Khalil$^b$, Stefano Moretti$^{c,d}$, Shoaib Munir$^a$, Roman Nevzorov$^e$, Alexandre Nikitenko$^f$, Harri Waltari$^d$} \\[0.25cm]
{\small \sl $^a$ East African Institute for Fundamental Research (ICTP-EAIFR),
University of Rwanda, Kigali, Rwanda} \\[0.25cm]
{\small \sl $^b$ Center for Fundamental Physics, Zewail City of Science and Technology, 6 October City, Giza, Egypt} \\ [0.25cm]
{\small \sl $^c$ School of Physics \& Astronomy,
University of Southampton, Southampton SO17 1BJ, UK} \\[0.25cm]
{\small \sl $^d$ Department of Physics and Astronomy,
Uppsala University, Box 516, SE-751 20 Uppsala, Sweden} \\ [0.25cm]
{\small \sl $^e$ I. E. Tamm Theory Department, P.N. Lebedev Physical Institute of the Russian Academy of Sciences, 53 Leninskiy Prospekt, 119991 Moscow, Russia} \\ [0.25cm]
{\small \sl $^f$ NRC Kurchatov Institute,
25 Bolshaya Cheremushkinskaya, 117218 Moscow, Russia} \\ [0.25cm]
{\small \url{mali@eaifr.org}, \url{skhalil@zewailcity.edu.eg}, \url{s.moretti@soton.ac.uk} \& \url{stefano.moretti@physics.uu.se}, \url{smunir@eaifr.org}, \url{nevzorovrb@lebedev.ru}, \url{Alexandre.Nikitenko@cern.ch}, \url{harri.waltari@physics.uu.se}}}
\end{center}
\vspace*{0.4cm}


\begin{abstract}
\noindent
We perform a model-independent search for leptoquarks (LQs) at the Large Hadron Collider through their pair-production and subsequent decay into $t\bar t \tau\tau$ intermediate states. We show that, assuming full luminosity of the Run 2, a fully hadronic signal emerging from this intermediate state can surpass in sensitivity the established searches relying on leptons in the final state. Our conclusion is supported by a thorough Monte-Carlo analysis, and we advocate the deployment of our proposed search channel in the proper experimental setting of the Run 3. Furthermore, in order to highlight the full scope of this approach for constraining LQ theories, we interpret our results in the context of the string-inspired Exceptional Supersymmetric Standard Model, which naturally predicts the $S_1$--type scalar LQ states that we analyse here. 
\end{abstract}


\newpage
\section{Introduction}
\label{sec:intro}

Leptoquarks are predicted by various extensions of the Standard Model (SM), such as Technicolour \cite{Eb:1981axe,PhysRevD.44.2678}, models of quark-lepton unification \cite{PhysRevD.10.275,FileviezPerez:2013zmv,Blanke:2018sro,Aydemir:2018cbb,Heeck:2018ntp}, and grand unification theories (GUTs) based on $SU(5)$, $SO(10)$, $E_6$~\cite{Das:2016vkr,Marzocca:2018wcf,Becirevic:2018afm,Aydemir:2019ynb,Crivellin:2019dwb,Aydemir:2022lrq}, and etc. They are rather unique amongst the many new particles predicted in new physics frameworks beyond the SM for being 
colour-triplet bosons that carry both lepton and baryon numbers. Their other quantum numbers (i.e., spin, (fractional) electric charge, and weak isospin) can be different in different theories.

Naturally, they were extensively searched for at the $ep$ colliders. In fact, at HERA, DESY, an excess was found in 1997 \cite{H1:1997utt} which would have been compatible with the so-called first-generation LQs (i.e., those coupling to first-generation leptons and quarks). However, further searches at HERA (and also at Tevatron, where most of the partonic processes were initiated by quarks inside the (anti-)proton) failed to confirm the anomaly. As a result, mass limits near 300 GeV \cite{PhysRevD.68.052004,H1:2011cqr} were set on first-generation LQs. Even weaker limits were placed on second-generation LQs. These (relatively) weak limits were owing to the fact that the partonic energy accessible at HERA and Tevatron was limited to the sub-TeV range. However, as colour triplets, LQs also interact with gluons. At the Large Hadron Collider (LHC), since the amount of gluons inside the proton can well exceed that of (both valence and sea) quarks, the availability of much higher centre-of-mass energy may enable the pair-production of third-generation LQs. 

The ATLAS and CMS experiments have both, therefore, analysed the datasets from the LHC Runs 1 and 2 to put exclusion limits on the mass of the third-generation LQs, based on their $t\bar t\tau\tau$, $t\bar b \tau\nu$, and $t\bar b \mu\nu$ decay channels, in the TeV range \cite{CMS:2018svy,CMS:2020wzx,ATLAS:2021oiz,ATLAS:2021jyv}. In this paper, we revisit the $t\bar{t}\tau \tau$ search channel for LQs of the $S_1$ type (i.e., LQs with charge --1/3 and spin zero) having couplings solely with the third generation fermions. However, unlike the existing searches for this channel, which have relied on leptons in the final state, we investigate the scope of the fully hadronic final state. Following a sophisticated detector-level analysis tensioning the signal to the most relevant backgrounds, we show that such a complementary channel can offer sensitivities to this specific type of LQ comparable to, if not better than, what has so far been achieved in their probes exploiting leptons. We then interpret the results of our analysis in the Exceptional Supersymmetric Standard Model (E$_6$SSM)~\cite{King:2005jy,King:2005my} (for a recent review, see Ref.~\cite{King:2020ldn}), as a theoretically well-motivated prototypical framework that naturally accommodates the $S_1$--type LQs.   

The plan of the paper is as follows. In the next section, we will describe our phenomenological setup in a model-independent way, which will then be subjected to a Monte-Carlo analysis in the following section. In Sec. \ref{sec:interp} we will discuss the E$_6$SSM and its possible scenarios that are amenable to experimental investigation by ATLAS and CMS, before concluding in Sec. \ref{sec:concl}.


\section{\label{sec:model} Third-generation Scalar LQs}

We consider a simple extension of the SM by adding one scalar $S_1$--type LQ, which is charged as $(3,1,-1/3)$ under its gauge group, $SU(3)_C\times SU(2)_L \times U(1)_Y$. We limit our discussion to LQs with the same quantum numbers as the coloured triplet of scalars, which, along with the Higgs doublet, constitutes the fundamental representation of $SU(5)$ \cite{FileviezPerez:2007bcw,Khalil:2013ixa}. LQs with such quantum numbers have been proposed as a solution to the $a_\mu$ and $B$-flavour anomalies (see, e.g., \cite{Cai:2017wry,Crivellin:2019qnh,Marzocca:2021azj}).

\begin{figure}[t]
    \centering
    \subfloat[]{{\includegraphics[scale=0.45]{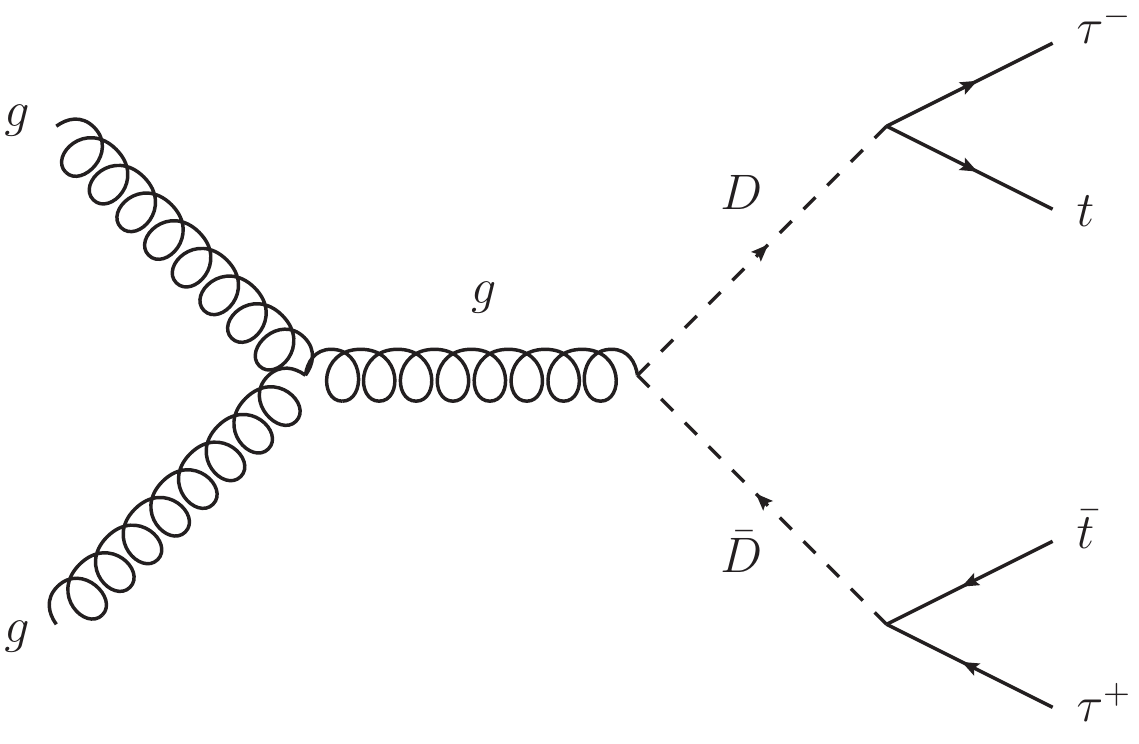} }}
    \qquad
    \subfloat[]{{\includegraphics[scale=0.45]{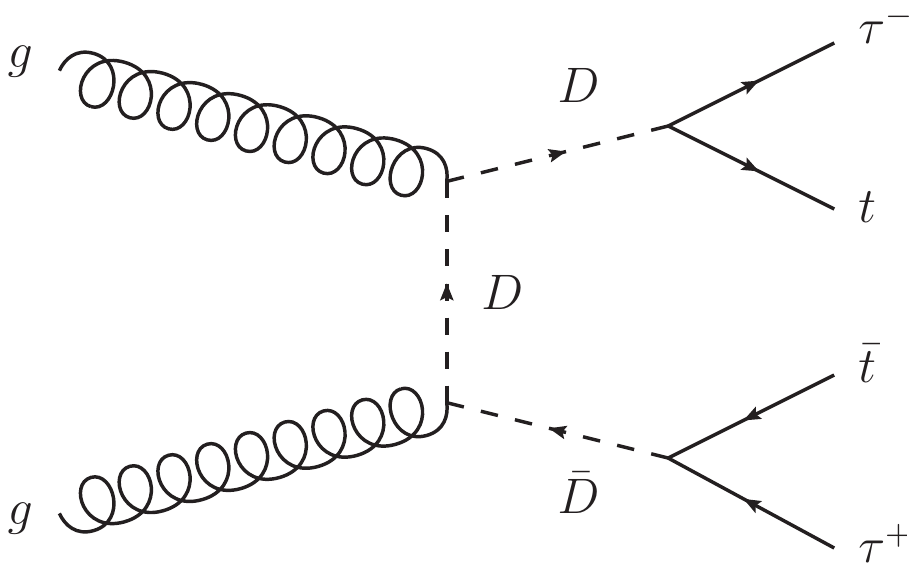} }}
    \qquad
    \subfloat[]{{\includegraphics[scale=0.45]{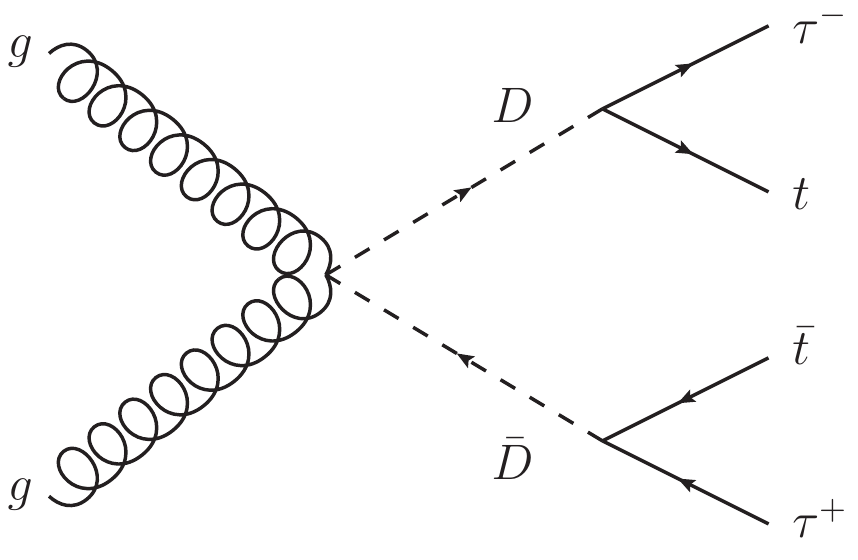} }}
    \qquad
    \subfloat[]{{\includegraphics[scale=0.45]{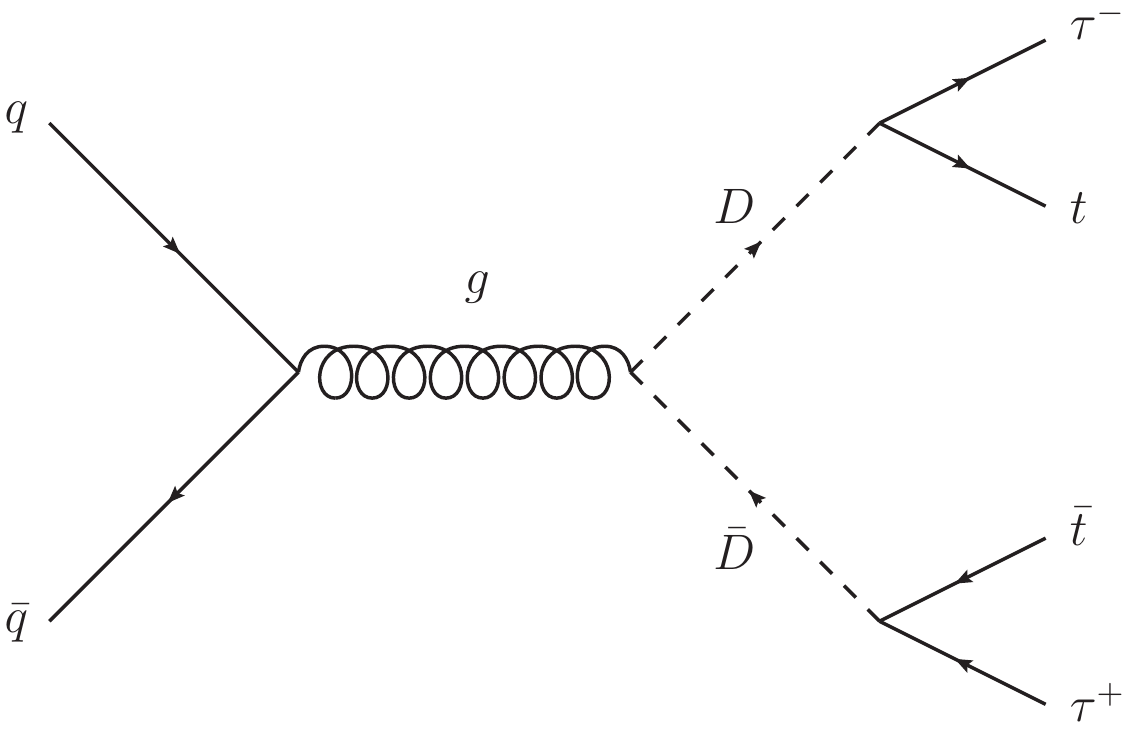} }}
    \qquad
    \subfloat[]{{\includegraphics[scale=0.45]{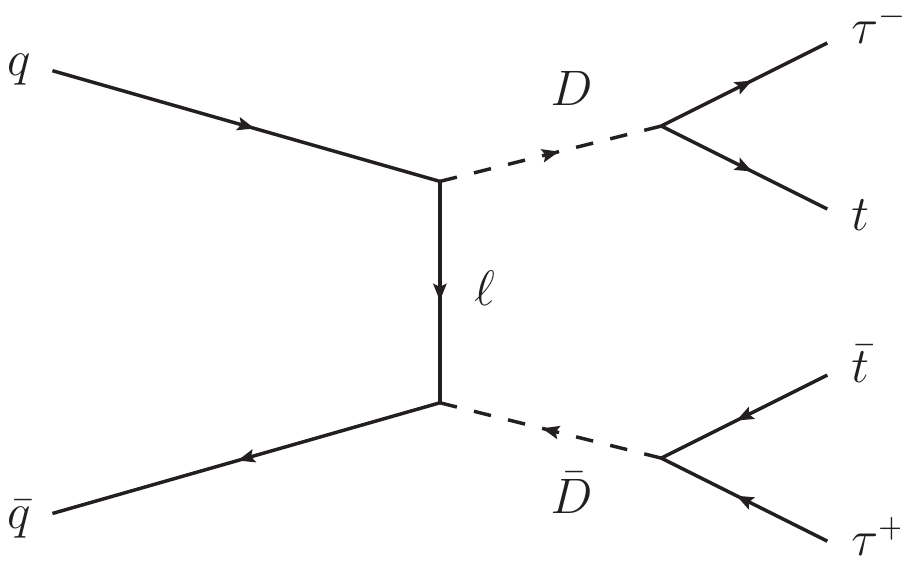} }}
    \caption{Pair-production of scalar LQs at the LHC.}
    \label{FD}
\end{figure}

We assume that our LQ, denoted by $D$ henceforth, couples only to the third-generation fermions, with the corresponding Lagrangian being
\be
{\cal L} = \lambda \bar{Q}_L D \bar{\ell}_L + \lambda' \bar{t}_R D \bar{\tau}_R\,,
\ee
where $Q_L = \left(t, b\right)_L^T$, and $\ell_L = ( \nu_\tau, \tau )_L^T$. This implies that the left-handed $t$ quark couples to the left-handed $\tau$ lepton, while the right-handed $t$ couples to the right-handed $\tau$. The mass of the $t$ quark at the one-loop level can thus cause the $\tau$ chirality to flip. Based on the above interaction terms, the $D$ can decay into $t\tau$ and $b\nu_\tau$ pairs. However, we assume $\lambda' \gg \lambda$, so that the decay to $t \tau$ is significantly enhanced. This LQ can be produced in pairs at the LHC through the processes shown in Fig.~\ref{FD}. 

Given that the $D$ couples only to the third-generation fermions, the diagram (e) in Figure \ref{FD} is negligible (unless the corresponding $\lambda$ and $\lambda^{\prime}$ couplings are extremely large, which is never the case here). The cross section for the $D$ pair production is therefore almost completely determined by QCD, and can be considered to be model-independent.\footnote{The contribution from the model-dependent diagram (e) can, in fact, be distinguished from the others as one expects two $b$--tagged jets close to the beam direction.} We can write this in the form
\be
\sigma(p p \to t\bar{t}\tau\tau) = \sigma(p p \to D \overline{D}) \times {\rm {\rm BR}}(D \to t  \tau^-) \times  {\rm {\rm BR}}(\overline{D} \to \bar{t} \tau^+)\,,
\ee
where BR stands for branching ratio. In our setup, we have ${\rm {\rm BR}}(D \to t \tau^-) =  {\rm {\rm BR}}(\overline{D} \to \bar{t}\tau^+)\simeq 1$, and the leading order (LO) partonic cross sections $ \sigma(g g \to D \overline{D})$ and $\sigma(q\bar{q}\to D\overline{D})$ are given by \cite{Kramer:1997hh}
\ba
 \sigma(g g \to D \overline{D}) & = & \frac{\alpha_s^2 \pi}{96 \hat{s}} \left[ \beta(41 -31 \beta^2) + (18 \beta^2 - \beta^4 -17) \log \frac{1+\beta}{1-\beta}\right]\,,~~{\rm and}\\
 \sigma(q\bar{q}\to D \overline{D}) & = & \frac{2\alpha_{s}^{2}\pi}{27\hat{s}}\beta^{3},
 \ea
 where $\beta= (1- 4 M_D^2/\hat{s})^{1/2}$, and $\sqrt{\hat{s}}$ is the partonic centre-of-mass energy.

In Tab. \ref{tb:xsecs} we show the LO and next-to-LO (NLO) pair-production cross sections, computed with \textsc{MadGraph5 v3.2.0} \cite{Alwall:2014hca} and \textsc{Prospino v2.1}\cite{Beenakker:1996ed}, using the NNPDF 3.1 parton distribution functions (PDFs) at LO \cite{NNPDF:2017mvq}. Note that our cross section values are somewhat smaller than those in \cite{ATLAS:2021oiz}, mainly due to the different PDF sets used, but they agree well with those in Ref. \cite{Borschensky:2020hot}, which also uses NNPDF 3.1.
\begin{table}\centering
\begin{tabular}{l|c c}
$m_{D}$ (GeV) & $\sigma_{\rm LO}$ (fb) & $\sigma_{\rm NLO}$ (fb)\\
\hline
$1000$ & $3.22$ & $5.73$\\
$1100$ & $1.57$ & $2.86$\\
$1200$ & $0.79$ & $1.50$\\
$1300$ & $0.42$ & $0.79$\\
$1400$ & $0.22$ & $0.44$\\
$1500$ & $0.12$ & $0.25$\\
$1600$ & $0.070$ & $0.146$
\end{tabular}
\caption{The LO and NLO production cross sections for the $pp\rightarrow D\overline{D}$ process for a range of the $D$ mass at the $\sqrt{s}=13$~TeV LHC. The uncertainty related to the variation of the renormalisation and factorisation scales is $\pm 20\%$. \label{tb:xsecs}}
\end{table}

Our simplified extension of the SM with a $S_1$--type LQ can be considered an effective low-energy limit of some GUT framework. In the simplest GUTs, however, our $D$ would have non-zero quark-lepton as well as quark-quark couplings, which can lead to rapid proton decay. To ensure the model's phenomenological viability, all the couplings of the $D$ to quark pairs can be set to zero, implying the conservation of baryon and lepton numbers. 

Furthermore, GUT models have a large separation of scales, and to stabilise the ensuing hierarchy, supersymmetry is often preferred. If we supersymmetrise our model, we need to introduce two chiral superfields, one of which has $\lambda$--type and the other $\lambda^{\prime}$--type superpotential couplings. The `exotic' particle spectrum then consists of two scalars and one fermion. The scalar states may have non-zero mixing, and one of the resulting physical states decays dominantly to $t\tau$, while the other has nearly equal BRs in the $t\tau$ and $b\nu$ decay channels. The decay of their superpartner fermion LQ leads to third-generation fermions and missing transverse momentum. Such a low-energy effective model can be embedded in a supersymmetric (SUSY) GUT framework such as the E$_6$SSM, which is free from the severe constraints from proton stability noted above.  

\section{\label{sec:fullyhadronic} The fully hadronic $t\bar{t}\tau\tau$ final state at the LHC Run 2}

\begin{table}[tbp]\centering
\begin{tabular}{|c|c|c|} \hline
Decay Mode & Mass Limit [GeV] & Experiment  \\ \hline
 \hline
\multirow{2}{*}{$t\bar{t}\tau\tau$} & 900 & CMS \cite{CMS:2018svy} \\
 & 1400 & ATLAS \cite{ATLAS:2021oiz} \\
 \hline
\multirow{3}{*}{$tb\tau\nu$} & 950 & CMS \cite{CMS:2020wzx} \\
 & 1250 & ATLAS \cite{ATLAS:2021jyv}\\
 & 1220 & ATLAS \cite{ATLAS:2021oiz}\\
 \hline
\end{tabular}
 \caption{\label{TAB:limits}$95\%$ CL lower mass limits by the ATLAS and CMS Collaborations from pair-produced third-generation scalar LQs.}
\end{table}

The $S_1$--type LQ under consideration here is subject to stringent limits from the ATLAS and CMS collaborations, most recent ones of which are given in Table~\ref{TAB:limits}. These limits correspond to the BR$(t\tau)$ at the 95\% confidence level (CL), and are obtained either directly when $t\tau$ is the exclusive decay channel of the LQ, or translated from the measurements in the $b\nu$ decay channel, when it is additionally open, as $1-{\rm BR}(b\nu)$. For the $t\tau$ decays of the pair-produced scalar LQs, these searches have so far employed final states with $e$ and $\mu$ leptons as well as hadrons from the $\tau$ decays ~\cite{ATLAS:2021oiz,CMS:2022nty}. For collider analyses of some other possible LQ decay channels, see \cite{Vignaroli:2018lpq,Ghosh:2022vpb} and references therein.

In this study, we propose to complement the ATLAS and CMS probes of the $t\bar{t}\tau\tau$ intermediate state with the fully hadronic topology emerging when the decays of both the $\tau$ leptons and both the $t$ quarks result in only jets in the final state. The analysis should select events with two $\tau_{\rm h}$ (i.e., $\tau$ decaying hadronically) and at least six jets, with two of them being $b$--tagged. The fully hadronic final state has two advantages compared with the semi-leptonic final state. First, it allows one to fully reconstruct the $D$ mass. Second, it increases the number of the signal events. We will show that the level of the background can be reduced to that of the expected signal using $\tau$-- and $b$--tagging techniques, and by exploiting the large mass of the $\tau_{\rm h}$ pair.

To investigate the scope of the fully hadronic final state, we performed a signal-to-background analysis for the pair-production of LQs with a mass of 1\,TeV each. The signal and the dominant $t \bar{t} j j$ background cross sections were calculated with Madgraph\_aMC@NLO at the LO, with selections $\Delta R(jj)>0.4$, $p_{T}^{j}>20$\,GeV, and $|\eta^{j}|<2.4$ for the jet transverse momentum and pseudorapidity, respectively, at the parton level. $\tau$ leptons in the signal sample were forced to decay hadronically, $\tau^{\pm}\rightarrow\pi^{\pm}\pi^{0}\nu_{\tau}$. We used the efficiencies and fake rates for $\tau$-- and $b$--tagging, and the $\tau\tau$ mass resolution for the $Z$ and Higgs bosons from the CMS publications (cited in the following), implying
\begin{itemize}
\item $p_{1(2)}^{\tau}=0.7$ (0.5) efficiency of $\tau$--tagging for real $\tau$, with
$p_{T}^{\tau_{\rm h}}>100$\,GeV ($<100$\,GeV), and $p_{\tau-{\rm fake}}=3\times10^{-3}$ mistagging rate for the jets (using the DeepTau algorithm~\cite{CMS:2022prd});
\item $p_{b}=0.8$ efficiency of $b$--tagging for real $b$--jets, and $p_{b-{\rm fake}}=10^{-2}$ mistagging rate for non--$b$--jets (using the DeepJet algorithm~\cite{Bols:2020bkb});
\item 15\% di--$\tau$ mass resolution for the $Z$ and Higgs bosons, i.e., 14\,GeV and 19\,GeV, respectively (using the SVFit algorithm~\cite{Bianchini:2014vza}).
\end{itemize}

\begin{figure}[tbp]
  \begin{center}
    \resizebox{7cm}{!}{\includegraphics{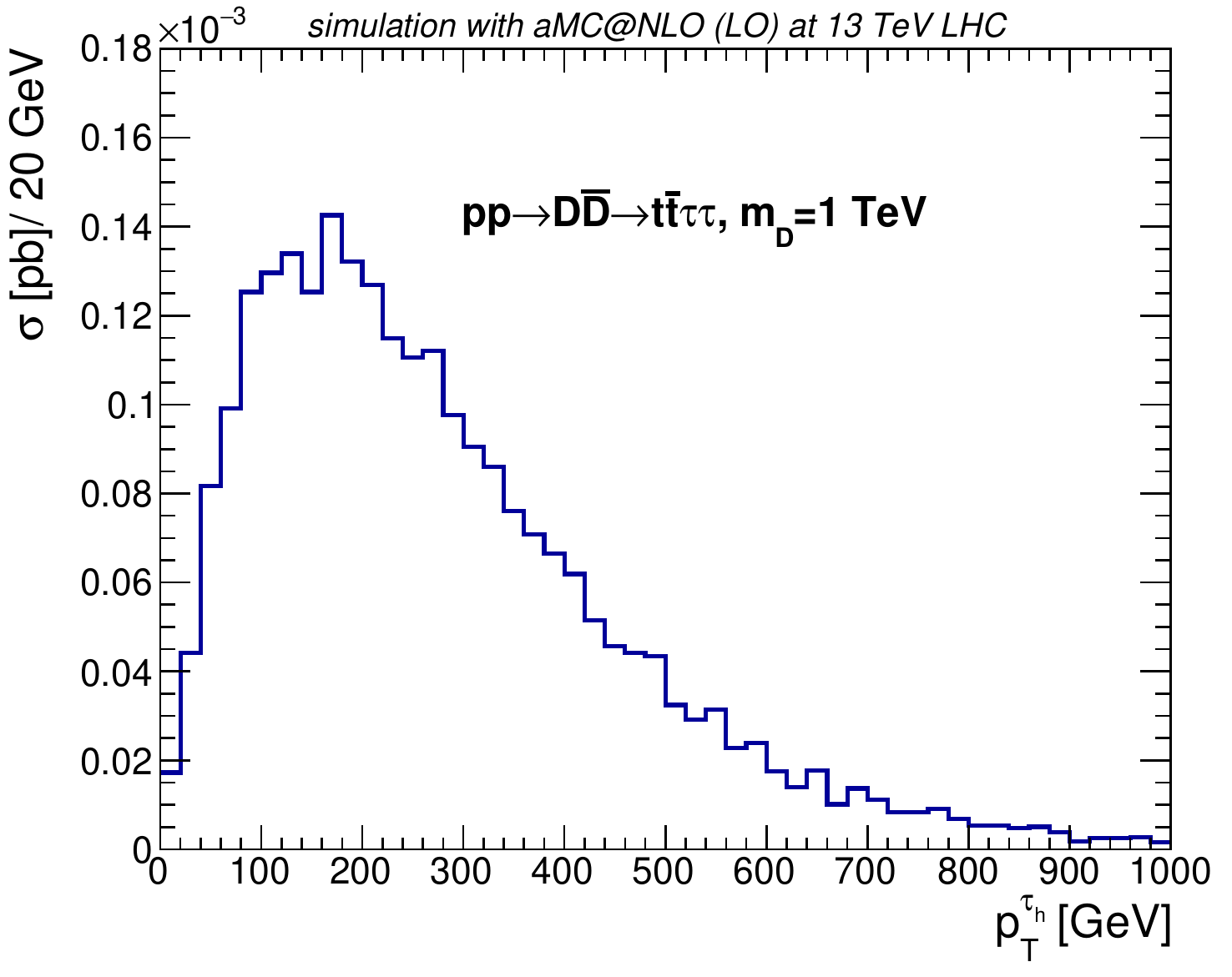}}
    \hfill
    \resizebox{7cm}{!}{\includegraphics{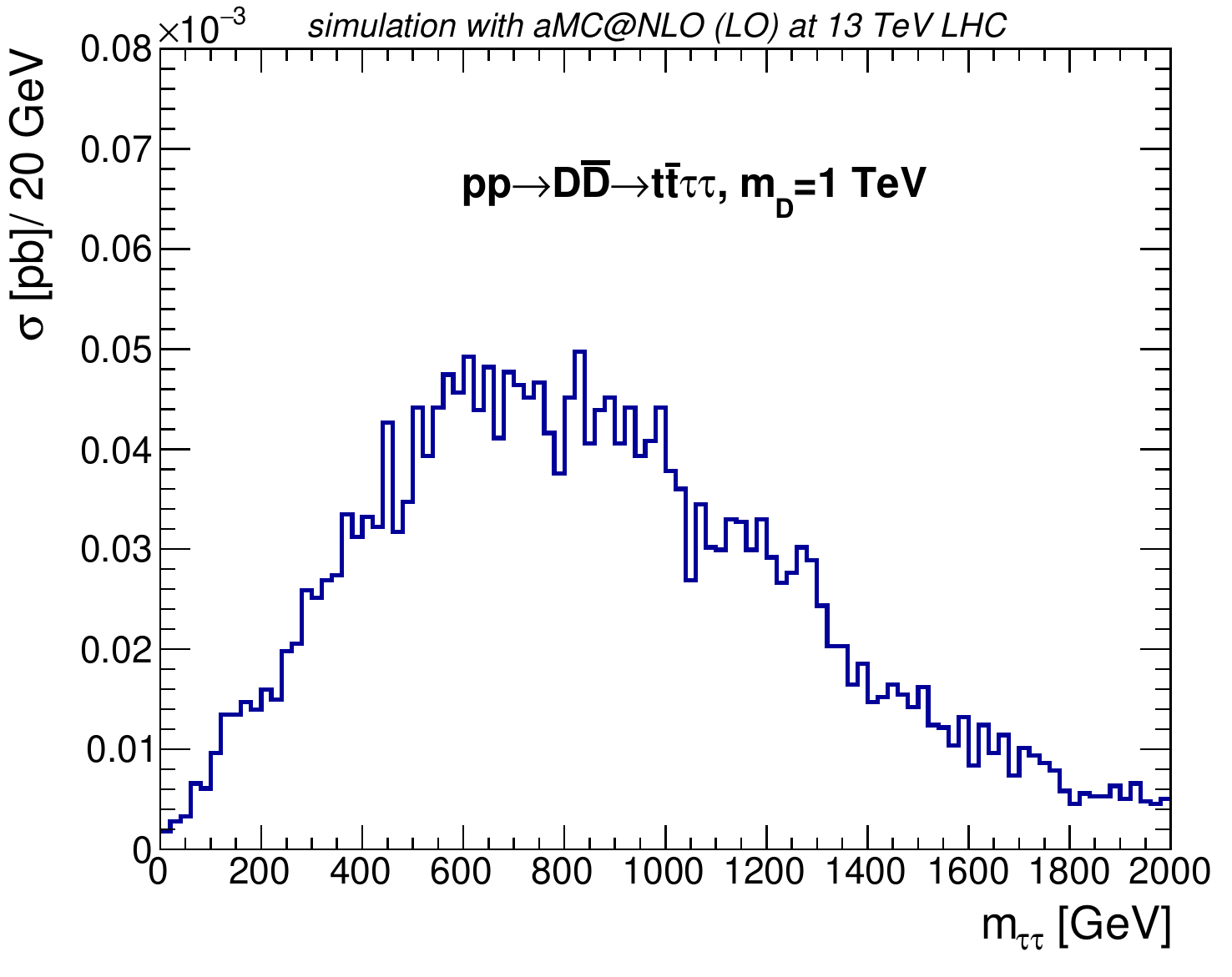}}
     \caption{$p_T$ of the $\tau_{\rm h}$ (left) and the di--tau effective mass (right) for the $D\overline{D}$ signal.}
    \label{fig:DD}
  \end{center}
\end{figure}

\begin{figure}[tbp]
  \begin{center}
    \resizebox{7cm}{!}{\includegraphics{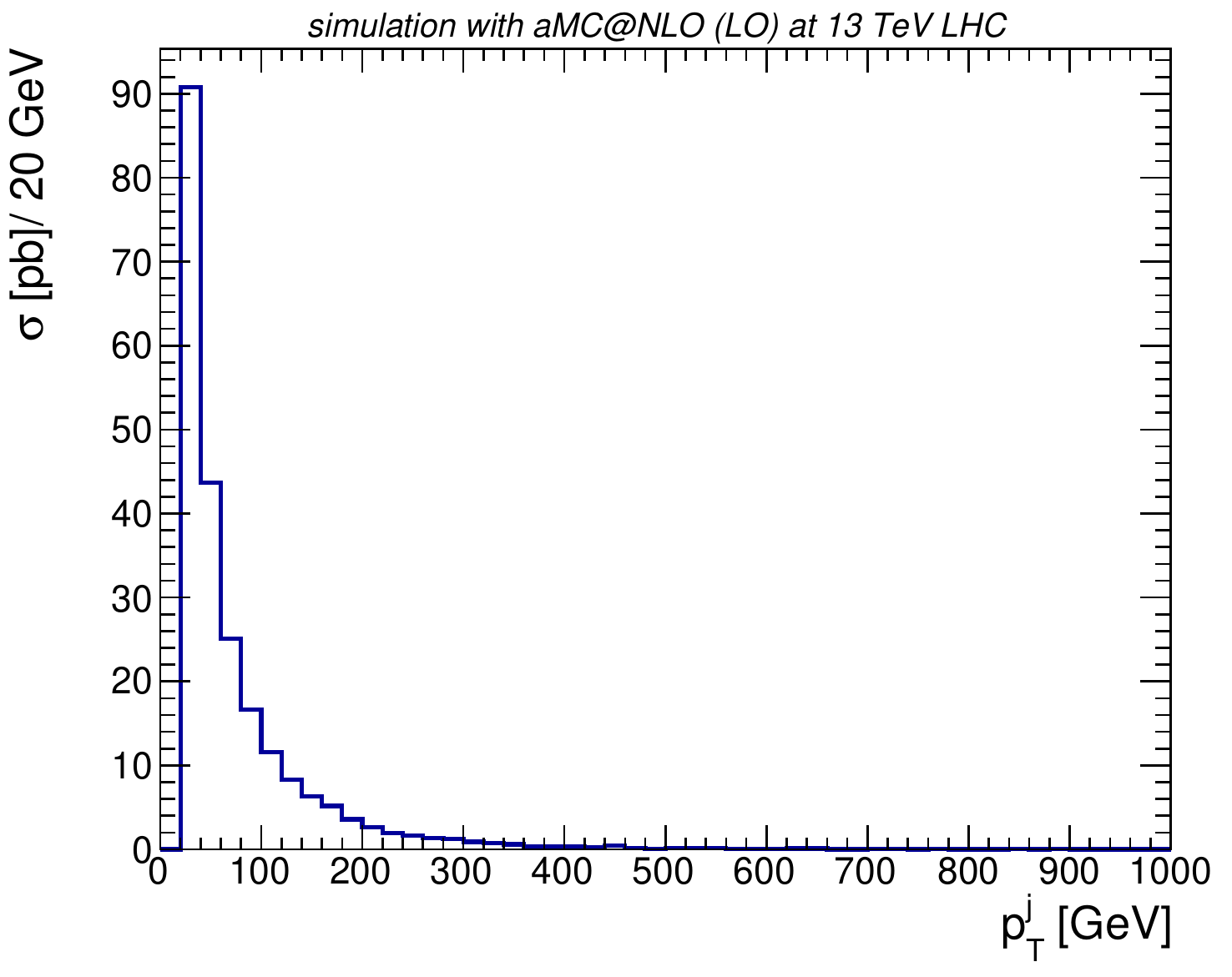}}
    \hfill
    \resizebox{7cm}{!}{\includegraphics{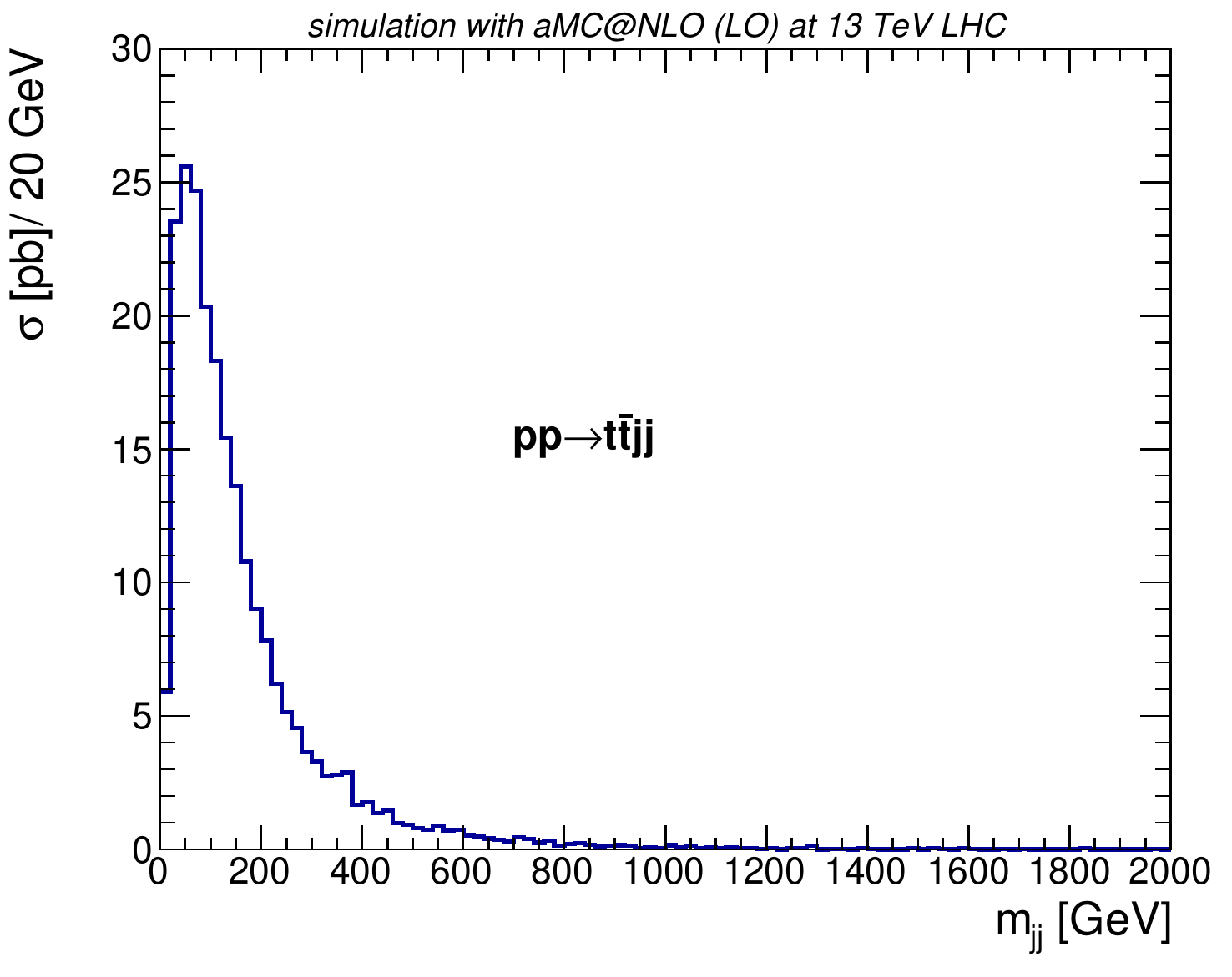}}
     \caption{$p_T$ of the parton jet (left) and the di--jet effective mass (right) for the $t\bar{t}jj$ background.}
    \label{fig:ttjj1}
  \end{center}
\end{figure}

Fig.~\ref{fig:DD} shows the distribution of $p_{T}^{\tau_{\rm h}}$
(left), and the $\tau\tau$ effective mass, $m_{\tau\tau}$, (right) for the signal. For the $t\bar{t}jj$ events, we show in the left panel of Fig.~\ref{fig:ttjj1} the distribution of the $p_T$ of the parton jets, i.e., the two partons in addition to $t\bar{t}$, and in the right panel the di--jet effective mass, $m_{jj}$, reconstructed from these parton jets.

We exploit the difference between the $m_{\tau\tau}$ and $m_{jj}$ distributions to suppress
the dominant $t\bar{t}jj$ background while preserving a high efficiency for the signal.
The event selection strategy we propose is the following. Events must contain eight jets with
two of them $b$--tagged. Six jets in the event, including the two $b$--tagged jets, must be associated
with the two $t$ quarks. The remaining two jets have to be $\tau$--tagged, and their effective mass should be greater than 200\,GeV. In our estimate of the efficiency of the di--jet mass
selection for the $t\bar{t}jj$ background we assume correct association of
the jets with the $t$ quarks for 100\% of the events. It was shown, however, in~\cite{CMS:2019eih} that the fraction of the $t\bar{t}jj$ events with correct $j$--to--$t$ association is 60\%. It could thus potentially make the suppression factor of the di--jet mass selection used in our
estimates weaker. 

For the estimation of the efficiency of the di--jet mass selection we used the generator-level $\tau_{\rm h}$ for the signal, and parton jets with $p_T>40$\,GeV and $|\eta|<2.4$ for the $t\bar{t}jj$ background. This corresponds to the CMS di--$\tau$ high-level trigger
selections used for the Run 2~\cite{CMS:2022goy}. Table~\ref{tab:bkgs} shows the cut-flow for the signal and background cross sections before and after the selections, along with the selection efficiencies. The selections considered here are the double--$b$-- and double--$\tau$--tagging, and the constraint on the di--$\tau$ mass. 

For the signal cross section we have used two values. The cross section corresponding to the upper limit obtained in the ATLAS analysis~\cite{ATLAS:2021oiz} at $m_{D}$=1\,TeV, and (in parenthesis) the cross section of the $pp \rightarrow  D\overline{D}$ process in the E$_6$SSM at the 13\,TeV LHC. The latter includes the NLO corrections calculated with Prospino~\cite{Beenakker:1996ed}. For the background, we have used the CMS measurements of the inclusive $t\bar{t}jj$ and $t\bar{t}b \bar{b}$ cross sections~\cite{CMS:2020grm,CMS:2019eih}, the inclusive cross sections for the $t\bar{t}Z(\rightarrow\tau\tau)$~\cite{CMS:2019too} and $t\bar{t}W (\rightarrow jj)$~\cite{CMS:2017ugv} processes, and the $Z(\rightarrow\tau\tau)+6j$~\cite{CMS:2018mdf} and $W(\rightarrow jj)+6j$~\cite{CMS:2017gbl} cross sections, as functions of the jet multiplicity. The cross section for the Higgs boson production, in association with two $t$ quarks, followed by its $\tau\tau$ decay was taken from Ref.~\cite{LHCHiggsCrossSectionWorkingGroup:2016ypw}. 

For two background processes, $t\bar{t}W$ and $W+6j$, we did not evaluate the efficiency of the di--$\tau$ mass selection, and therefore give the upper limit on the corresponding cross sections after the selections. Note that the di--$\tau$ mass selection cuts eliminates the $t\bar{t}Z$ and $Z+6j$ backgrounds due to more than 5$\sigma$ difference between the di--jet threshold and $m_Z$. It likewise eliminates the $t\bar{t}H$ background, since the $jj$ threshold is also about 4$\sigma$ above the $m_H$.

\begin{table}[tbp]
\begin{center}
{\footnotesize
\begin{tabular}{|c|c|c|c|c|c|} \hline
Process & $\sigma$ [fb] & $bb$--tagging & $\tau\tau$--tagging & $m_{\tau\tau}>200$\,GeV & $\sigma^{\rm sel}$[fb]\\ \hline\hline
Signal & 0.5 (5.73) & $p_{b}^{2}$ & $(p_{1}^{\tau})^{2}$ & 0.97 & 0.15 (1.74) \\
\hline\hline
$t\bar{t}jj$ & $275\times 10^{3}$& $p_{b}^{2}$ & $(p_{\tau-{\rm fake}})^{2}$ & 0.45 & 0.71 \\
\hline
$t\bar{t}Z$ & $950\times{\rm BR}(Z\rightarrow\tau\tau)$ & $p_{b}^{2}$ & $(p_{2}^{\tau})^{2}$ & $<5.7\times10^{-7}$ & $<2.6\times 10^{-6}$ \\
\hline
$t\bar{t}W$ & $770\times{\rm BR}(W\rightarrow q q^\prime)$ & $p_{b}^{2}$ & $(p_{\tau-{\rm fake}})^{2}$ &  & $<3.1\times 10^{-3}$ \\
\hline
$t\bar{t}H$ & 32 & $p_{b}^{2}$ & $(p_{2}^{\tau})^{2}$ & $6.3\times 10^{-5}$ & $3.2\times 10^{-4}$ \\
\hline
$Z+6j$ & 50 & ${\rm C}_{6}^{2}p_{b-{\rm fake}}^{2}(1- p_{b-{\rm fake}})^{4}$ & $(p_{2}^{\tau})^{2}$ & $<5.7\times 10^{-7}$ & $<1.0\times 10^{-8}$ \\
\hline
$W+6j$ & 600$\times\rm R^{q q^\prime}_{\mu\nu}$ & ${\rm C}_{8}^{2}p_{b-{\rm fake}}^{2}(1- p_{b-{\rm fake}})^{6}$ & $(p_{\tau-{\rm fake}})^{2}$ & & $<1.0\times 10^{-4}$ \\
\hline
\end{tabular}
}
\caption{The signal and background cross sections before and after the selections, and the selection efficiencies. The $C^{k}_{n}$ are the binomial coefficients $\binom{n}{k}$. See text for details. \label{tab:bkgs} }
\end{center}
\end{table}

\begin{figure}[tbp]
  \begin{center}
    \resizebox{7cm}{!}{\includegraphics{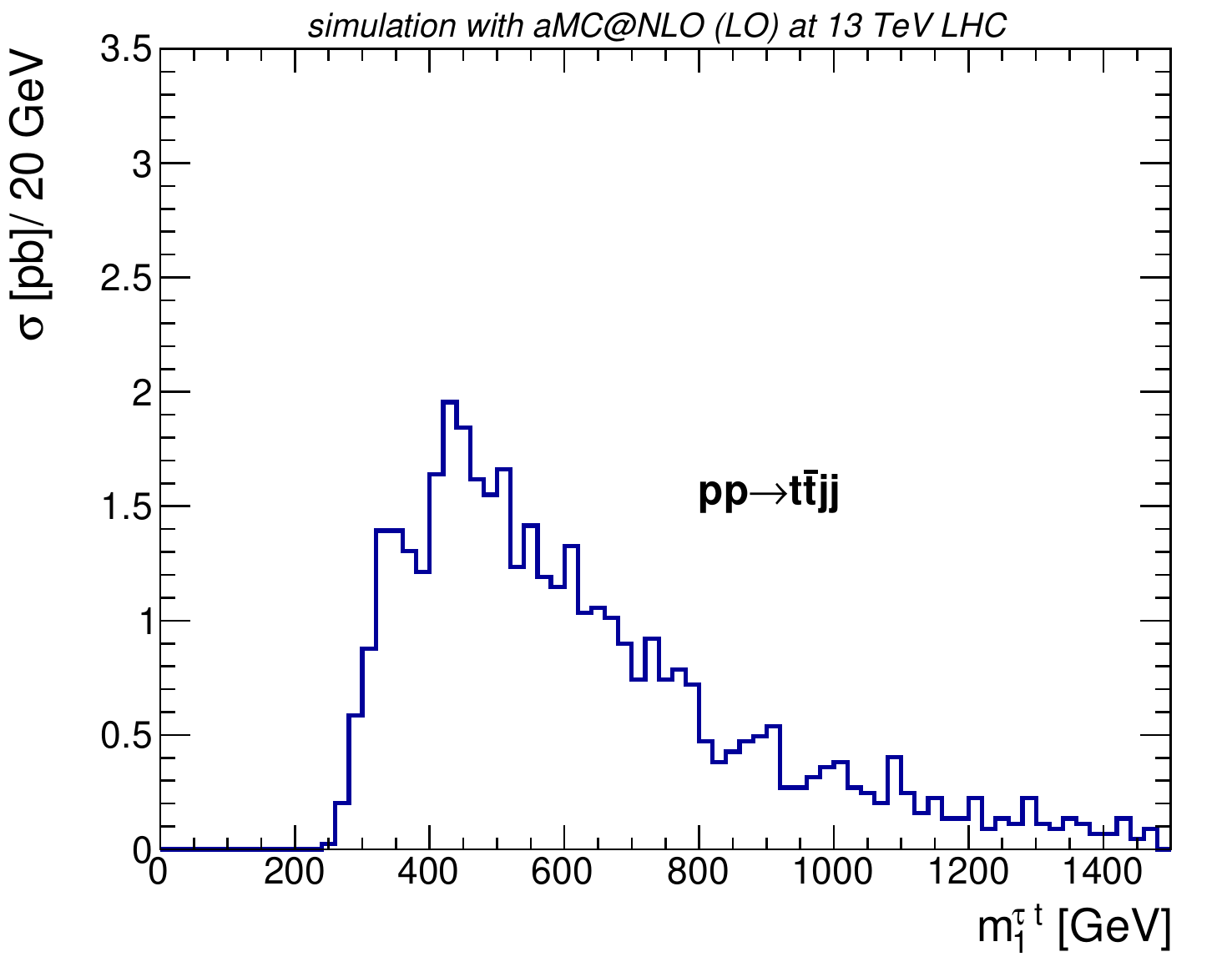}}
    \hfill
    \resizebox{7cm}{!}{\includegraphics{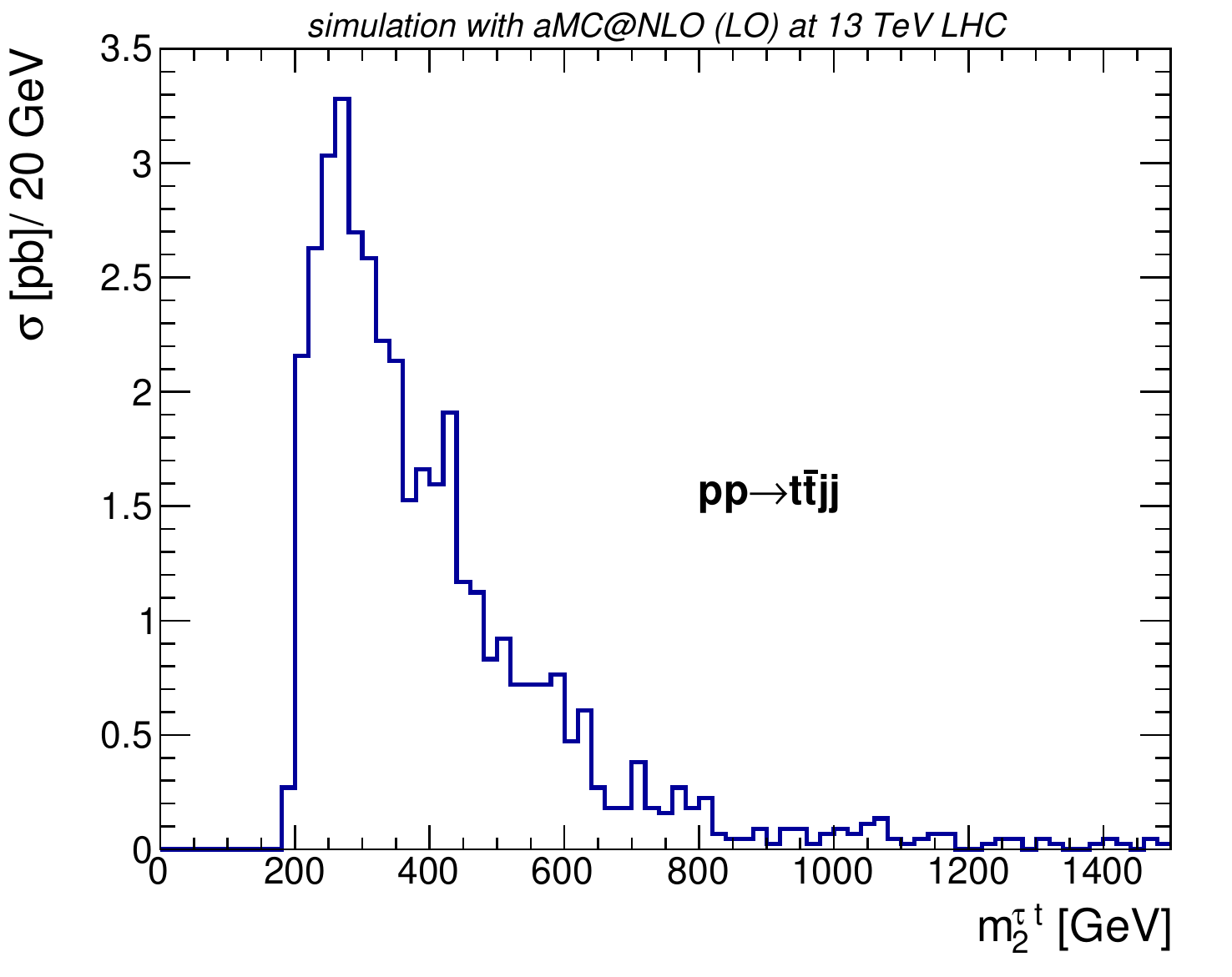}}
     \caption{Distribution of the $m_{1}^{t\tau}$ (left) and
              $m_{2}^{t\tau}$ (right) for the $t\bar{t}jj$ events
              after the selections $p_{T}^{j}>40$\,GeV, $|\eta^{j}|<2.4$, and $m_{jj}>200$\,GeV.}
    \label{fig:ttjj2}
  \end{center}
\end{figure}

As one can see from Table~\ref{tab:bkgs}, the selection cuts used eliminate all the backgrounds except
the dominant $t\bar{t}jj$. Further suppression of this background can be done by exploiting the possibility of the full reconstruction of the $D$ mass as the effective mass of the $t\tau$ pair. 
The association of the two $\tau$'s and the two $t$ quarks to the $D\overline{D}$ pair can be done as follows. Out of the two $\tau_{\rm h}$ objects, we take the one which gives the $t\tau$ mass, $\rm m_{1}^{t\tau}$, closest to the $D$ mass (1\,TeV here). At this step, the $t$ quark out of the
two is taken randomly. The other $\tau_{\rm h}$ and the other $t$ quark then give the mass of the second LQ, $m_{2}^{t\tau}$. Fig.~\ref{fig:ttjj2} shows the distribution of the $m_{1}^{t\tau}$ (left) and $m_{2}^{t\tau}$ (right) for the $t\bar{t}jj$ events after the selections
$p_{T}^{j}>40$\,GeV, $|\eta^{j}|<2.4$, and $m_{jj}>200$\,GeV.

One can see that selecting a large $m_{2}^{t\tau}$ can sufficiently suppress the
$t\bar{t}jj$ background. The detector mass resolution of the $t$ quark in
a fully hadronic decay mode is 14\,GeV (8\% of the $m_t$) at the CMS~\cite{CMS:2018tye}.
We assume conservatively a detector resolution of 10\% for the $t\tau$ mass, so that $\sigma_D^{\rm exp}=100$\,GeV for $m_{D}=1$\,TeV, and the natural width of $D$ to be negligible compared to it. The efficiency of the selection $m_{2}^{t\tau}>m_{D}-2\sigma_{D}^{\rm exp}=800$\,GeV is 0.06 for the $t\bar{t}jj$ background, and 0.95 for the signal. It leads to cross sections of 0.14 (1.65) fb for the signal and 0.04 fb for the $t\bar{t}jj$ background after all the selections imposed so far. It corresponds to 20 (231) expected signal events and 6 background events for an integrated luminosity of 140 fb$^{-1}$, resulting in a signal significance, $2(\sqrt{S+B}-\sqrt{B})$~\cite{Bityukov:1998ju}, of 5.2 ($>5$). Moreover, the expected upper limit on the number of the signal events is 6.2 corresponding to the cross section of 0.16 fb, which is better than the 0.5 fb limit at $m_{S_1}=1$\,TeV from the ATLAS analysis~\cite{ATLAS:2021oiz}.

Two factors related to the detector reconstruction effects can potentially weaken the performance of our analysed channel. First, as mentioned already, the association of the jets and $b$--jets to the $t$ quarks is not expected to be 100\% correct. And second, the jet energy smearing can reduce the impact of the $m_{jj}>200$\,GeV selection cut. Such effects can be estimated with a full detector simulation and reconstruction. We, however, do not expect our reported results to change drastically, and emphasise that the fully hadronic $t\bar{t} \tau \tau$ final state can nicely complement the current Run 2 analyses employing final states with leptons.

\section{\label{sec:interp} Interpretation in the E$_6$SSM}

As an application of our analysis to a top-down framework, we look at how it constrains the parameter space of the E$_6$SSM, which is one of the best-motivated $E_6$-inspired SUSY models. The simple extension of the SM discussed in section \ref{sec:model} may be embedded in this model, with the sparticle mass scale lying in the few TeV range. 

In the E$_6$SSM, the $E_6$ group is broken down to the SM gauge group plus an extra $U(1)_{N}$ symmetry around the GUT scale $M_X$, where
\begin{equation}
U(1)_N=\dfrac{1}{4} U(1)_{\chi}+\dfrac{\sqrt{15}}{4} U(1)_{\psi}\,.
\label{401}
\end{equation}
The $U(1)_\psi$ and $U(1)_\chi$ symmetries are associated with the subgroups
$E_6 \supset SO(10) \times U(1)_\psi \supset SU(5) \times U(1)_\chi \times U(1)_\psi$. It is assumed
that in the E$_6$SSM the matter parity $Z_{2}^{M}=(-1)^{3(B-L)}$ is also preserved.

In this SUSY model the anomalies are automatically cancelled if the low-energy particle spectrum includes three complete $27$-dimensional representations of $E_6$ ($27_i$ with $i=1,2,3$). Each 27-plet contains one generation of ordinary matter, a SM singlet field $\Phi_i$ carrying non--zero $U(1)_{N}$ charge, up- and down-type Higgs doublets $H^{u}_{i}$ and $H^{d}_{i}$, as well as exotic scalar quarks (squarks) $D_i$ and $\overline{D}_i$ with electric charge $\pm 1/3$. In order to avoid rapid proton decay via these exotic states, one must impose some additional symmetry. The simplest options are a $Z_2^L$ symmetry, which implies that all supermultiplets except the lepton ones are even, or a $Z_2^B$ symmetry, under which exotic quark and lepton supermultiples are odd, whereas the others remain even~\cite{King:2005jy,King:2005my}. In the first case the $D_i$ and $\overline{D}_i$ are diquarks (Model I). In the second scenario the exotic squarks carry baryon ($B_{D}=1/3$ and $B_{\overline{D}}=-1/3$) and lepton ($L_{D}=1$ and $L_{\overline{D}}=-1$) numbers simultaneously so that they are LQs (Model II).

The exotic states in the E$_6$SSM may also give rise to flavour changing processes. In particular, the scalar components of the supermultiplets $H^u_{i}$ and $H^d_{i}$ can interact with the SM fermions of different generations, thus contributing to the amplitude of $K^0$--$\overline{K}^0$ oscillations, and resulting in new channels of muon decay, such as $\mu\to e^{-}e^{+}e^{-}$. The non-diagonal flavour transitions can be suppressed by imposing $Z^{H}_2$ symmetry, under which all the matter supermultiplets except one SM singlet superfield ($\Phi\equiv \Phi_3$) and one pair of $H^u_{i}$ and $H^d_{i}$ (say $H_d\equiv H^d_{3}$ and $H_u\equiv H^u_{3}$) are odd~\cite{King:2005my,King:2005jy}. The discrete $Z^{H}_2$ symmetry can only be an approximate one because it forbids all operators that allow the lightest exotic quark or squark to decay.

The appropriate suppression of the flavour-changing processes can be achieved when all $Z^{H}_2$ symmetry-violating couplings are less than $10^{-4}$. Within this original E$_6$SSM, only $H_u$, $H_d$, and $\Phi$ form the Higgs sector, whereas the Higgs-like doublets $H^d_{\alpha}$ and $H^u_{\alpha}$ as well as the SM singlets $\Phi_{\alpha}$ ($\alpha=1,2$) do not acquire
vacuum expectation values (VEVs). The VEV of $\Phi=\varphi/\sqrt{2}$ breaks $U(1)_{N}$ gauge symmetry, generating the mass of the $Z'$ boson,
\begin{equation}
m_{Z'}\approx g^{\prime}_1 \tilde{Q}_{\Phi} \,\varphi\,\,,
\label{4001}
\end{equation}
and the effective $\mu$--term. In Eq.~(\ref{4001}) $g^{\prime}_1$ and $\tilde{Q}_{\Phi} = \dfrac{5}{\sqrt{40}}$ are the $U(1)_{N}$ gauge coupling and the $U(1)_{N}$ charge of the superfield $\Phi$, respectively. The LHC constraints require the extra $U(1)_{N}$ gauge boson to be heavier than
$4.5\,\mbox{TeV}$~\cite{CMS:2021ctt,ATLAS:2019erb}. To satisfy this constraint $\varphi$ should be larger than 12\,TeV.

The conservation of the $Z_{2}^{M}$ symmetry and $R$-parity in the E$_6$SSM ensures that the lightest SUSY particle (LSP) is stable. In this case the LSP and the next-to-LSP (NLSP) are mostly composed of the fermion components of the superfields $\Phi_{\alpha}$. Using the approach discussed in~\cite{Hesselbach:2007te,Hesselbach:2007ta,Hesselbach:2008vt},
it was shown that the masses of the LSP and NLSP ($\tilde{H}^0_1$ and $\tilde{H}^0_2$) are smaller than 60--65\,GeV~\cite{Hall:2010ix}. The couplings of these fermions to the SM particles tend to be rather small. Nevertheless if the LSP had a mass close to $m_Z/2$,
it could account for some of the observed cold dark matter (DM) relic density~\cite{Hall:2010ix}.
In these scenarios, the lightest Higgs boson decays mainly into either $\tilde{H}^0_1$ or $\tilde{H}^0_2$, while all other {\rm BR}s are highly suppressed. Such scenarios have been already excluded by the LHC experiments. When the LSP and NLSP are considerably lighter than $m_Z$, the annihilation cross section of $\tilde{H}^0_1\tilde{H}^0_1\to \mbox{SM particles}$
becomes too small, resulting in a DM density which is substantially larger than its measured value.

Since 2006 several modifications of the E$_6$SSM have been explored
\cite{King:2005jy,King:2005my,Nevzorov:2012hs,Athron:2014pua,Howl:2007hq,Howl:2007zi,Howl:2008xz,Howl:2009ds, Athron:2010zz,Hall:2011zq,Callaghan:2012rv,Callaghan:2013kaa,King:2016wep,Khalil:2020syr,Nevzorov:2022zns}. The implications of the $U(1)_{N}$ extensions of the MSSM have been studied for the $Z$--$Z'$ mixing~\cite{Suematsu:1997au}, the neutralino sector~\cite{Suematsu:1997au,Keith:1997zb,Keith:1996fv}, leptogenesis~\cite{Hambye:2000bn,King:2008qb,Nevzorov:2017gir,Nevzorov:2018leq} and electroweak symmetry breaking~\cite{Keith:1997zb,Suematsu:1994qm,Daikoku:2000ep},
the renormalisation group (RG) flow of couplings~\cite{Keith:1997zb,King:2007uj},
the renormalisation of VEVs~\cite{Sperling:2013eva,Sperling:2013xqa}, non-standard neutrino models~\cite{Ma:1995xk}, the DM~\cite{Hall:2010ix,Khalil:2020syr,Nevzorov:2022zns,Hall:2009aj}, and the signatures associated with the inert neutralino states~\cite{Khalil:2021afa,Khalil:2021tpz}.
Within the E$_6$SSM the upper bound on the lightest Higgs mass near the quasi-fixed point was examined
in~\cite{Nevzorov:2013ixa,Nevzorov:2015iya}. The corresponding quasi-fixed point is an intersection of
the invariant and quasi-fixed lines~\cite{Nevzorov:2001vj,Nevzorov:2002ub}.
The particle spectrum in the constrained E$_6$SSM (cE$_6$SSM) and its modifications has been analysed
in~\cite{Athron:2015vxg,Athron:2016gor,Athron:2011wu,Athron:2008np,Athron:2009ue,Athron:2009bs,Athron:2012sq}. The degree of the fine tuning and the threshold corrections were explored in~\cite{Athron:2013ipa,Athron:2015tsa} and \cite{Athron:2012pw}, respectively. 

The exotic matter in the E$_6$SSM may result in distinctive LHC signatures~\cite{King:2005jy,King:2005my,Howl:2007zi,Athron:2010zz,Athron:2011wu,King:2006vu,King:2006rh,Belyaev:2012si,Belyaev:2012jz}, and can give rise to non-standard Higgs decays~\cite{Hall:2010ix,Athron:2014pua,Nevzorov:2015iya,Nevzorov:2013tta,Hall:2010ny,Hall:2011au,Hall:2013bua,Nevzorov:2014sha,Athron:2016usd,Nevzorov:2020jdq}. In the E$_6$SSM with approximate $Z_2^H$ symmetry, one can also impose an exact $Z^S_2$ symmetry, which implies that only components of the $\Phi_{\alpha}$ superfields are odd~\cite{Hall:2011zq}. In this case the LSP and NLSP become massless and decouple. The presence of such massless fermions does not affect the Big Bang Nucleosynthesis if the $Z'$ boson is rather heavy~\cite{Hall:2011zq}. In this variant of the model, one of the lightest $R$-parity odd state is stable and may account for some of the observed DM matter relic abundance. Neglecting non-renormalisable interactions and all the $Z^{H}_2$ symmetry-violating couplings, the superpotential of this model can be written as
\begin{equation}
\begin{array}{rcl}
W_{\rm{E}_6\rm{SSM}} & = &  \lambda \Phi (H_u H_d) + \lambda_{\alpha} \Phi (H^d_{\alpha} H^u_{\alpha}) + \kappa_{i} \Phi (D_{i} \overline{D}_{i}) 
\\
& + & \dfrac{1}{2}M_{i}N_i^c N_i^c + h_{ij} N_i^c (H_u L_j) + W_{\rm MSSM}(\mu=0)\,,
\end{array}
\label{402}
\end{equation}
where $N^c_i$ and $L_j$ are supermultiplets of the right--handed neutrinos and left--handed leptons.

In the case of Model II, the $Z^{H}_2$ symmetry-violating terms, that permit the lightest exotic quark or squark to decay, can be presented in the following form
\begin{equation}
W_2 = g_{ijk} e^c_i u^c_j D_k + \overline{g}_{ijk} (Q_i L_j) \overline{D}_k\ + g^{\prime}_{ijk}d^{c}_{i}N^{c}_{j}D_{k}\,.
\label{403}
\end{equation}
Here $e^c_i$, $u^c_k$, $d^{c}_{i}$, and $Q_i$ are supermultiplets of the right-handed charged leptons, the right-handed up-type quarks, the right-handed down-type quarks, and the left-handed quark doublets, respectively.

The VEV of the $\Phi$ induces the masses of the exotic fermions, which in the leading approximation
are given by
\begin{equation}
\mu_{D_i}=\dfrac{\kappa_i}{\sqrt{2}}\,\varphi\,,\qquad\qquad
\mu_{\tilde{H}_{\alpha}}=\dfrac{\lambda_{\alpha}}{\sqrt{2}}\,\varphi\,,
\label{404}
\end{equation}
where $\mu_{D_i}$ are the exotic quark masses, while $\mu_{\tilde{H}_{\alpha}}$ are
the masses of the states composed of the fermionic components of the $H^d_{\alpha}$ and $H^u_{\alpha}$. These masses are determined by the values of the Yukawa couplings $\kappa_i$ and $\lambda_{\alpha}$.

The breakdown of the gauge symmetry in the E$_6$SSM may give rise to a substantial
mixing between the scalar components of the supermultiplets $D_i$ and $\overline{D}_i$.
Since we choose the field basis such that the Yukawa couplings of the $D_i$ and $\overline{D}_i$
to $\Phi$ are flavour diagonal, which leads to a mixing only between the exotic squarks
from the same family. As a consequence, the calculation of the exotic squark masses reduces
to the diagonalisation of three $2\times 2$ matrices
\begin{equation}
\begin{array}{c}
M^2(i)=\left(
\begin{array}{cc}
M_{11}^2(i) + \Delta_{11}(i) & \mu_{D_i} X_{D_i} + \Delta_{12}(i) \\
\mu_{D_i} X_{D_i} + \Delta_{12}(i) & M_{22}^2(i) + \Delta_{22}(i)
\end{array}
\right)\,,\\
\\
{\rm with}\quad M_{11}^2(i)=m^2_{D_i}+\mu_{D_i}^2+ \Delta_{D}\,,\quad M_{22}^2(i)=m^2_{\overline{D}_i}+\mu_{D_i}^2+\Delta_{\overline{D}}\,,\\
X_{D_i}=A_{\kappa_i}-\dfrac{\lambda}{\sqrt{2}\varphi}v_1 v_2\,,\quad{\rm and}\quad
\Delta_{\phi}=\dfrac{g^{'2}_1}{2} \biggl(\tilde{Q}_{\overline{D}} v_1^2 +
\tilde{Q}_D v_2^2 + \tilde{Q}_{\Phi} \varphi^2\biggr) \tilde{Q}_{\phi}\,,
\end{array}
\label{405}
\end{equation}
where $i=1,2,3$, $v_1$ and $v_2$ are the VEVs of the Higgs doublets $H_d$ and $H_u$,
$\tilde{Q}_{D} = -\dfrac{2}{\sqrt{40}}$ and $\tilde{Q}_{\overline{D}} = -\dfrac{3}{\sqrt{40}}$ are the $U(1)_{N}$ charges of the $D_i$ and $\overline{D}_i$, while $\Delta_{lm}(i)$ ($l,m=1,2$) are the contributions of loop corrections. $m^2_{D_i}$ and $m^2_{\overline{D}_i}$ in the above equation are the soft scalar masses of the $D_i$ and $\overline{D}_i$, whereas
$A_{\kappa_i}$ are the trilinear scalar couplings associated with the Yukawa couplings $\kappa_i$.
The parameters $m^2_{D_i}$, $m^2_{\overline{D}_i}$, and $A_{\kappa_i}$ break global SUSY.

The $U(1)_N$ $D$--term contributions to the masses of the exotic squarks are set by $m_{Z'}^2$, i.e.,
\begin{equation}
\Delta_{D}\approx -\dfrac{1}{5}m_{Z'}^2\,,\qquad
\Delta_{\overline{D}}\approx -\dfrac{3}{10}m_{Z'}^2\,.
\label{406}
\end{equation}
These contributions are negative because the $U(1)_N$ charge of the $\Phi$ and the $U(1)_N$ charges of the $D_i$ and $\overline{D}_i$ are opposite. On the other hand, the $U(1)_N$ $D$--term contributions to the masses of the ordinary squarks and sleptons are positive. As a result, in some parts of the E$_6$SSM parameter space the exotic squarks tend to be substantially lighter than the superpartners of the SM fermions.

The magnitude of mixing in the exotic squark sector is governed by the mixing parameters $X_{D_i}$
as well as the masses $\mu_{D_i}$. If $\mu_{D_i}$ are sufficiently large, the mixing effects can be so substantial that the corresponding lightest exotic squarks may be among the lightest SUSY particles in the E$_6$SSM. In the Model II, the lightest scalar LQ, which we denote by $D$ in order to identify it with the $S_1$--type LQ in our low-energy SM extension discussed earlier, is a linear superposition of the scalar components of the $D_1$ and $\overline{D}_1$ supermultiplets. Its mass is given by
\begin{equation}
\begin{array}{rcl}
m^2_{S_1}&=&\dfrac{1}{2}\Biggl[M_{11}^2(1) + \Delta_{11}(1) +  M_{22}^2(1) + \Delta_{22}(1)\\[4mm]
&-&\sqrt{\Biggl(M_{11}^2(1) + \Delta_{11}(1) - M_{22}^2(1) - \Delta_{22}(1)\Biggr)^2 + 4 \Biggl(\mu_{D_1} X_{D_1} + \Delta_{12}(1)\Biggr)^2} \Biggr]\,.
\end{array}
\label{407}
\end{equation}
To simplify our analysis of the $D$ in the E$_6$SSM, we set $A_{\kappa_1}=0$. Because of this the mixing between the scalar components of the $D_1$ and $\overline{D}_1$ is almost always small unless $M_{11}^2(1)\approx M_{22}^2(1)$. In this case, $D$ is almost entirely the scalar component of either $D_1$ or $\overline{D}_1$, and $m^2_{D}$ is defined by the smallest diagonal element of the mass matrix in Eq.~(\ref{405}), i.e., either by $M_{11}^2(1)$ or $M_{22}^2(1)$. Moreover, we focus on the part of the E$_6$SSM parameter space where $m_{D}$ is smaller than the mass of the lightest fermion LQ. Therefore, the decay modes of the $D$ are entirely determined by the Yukawa interactions in Eq.~(\ref{403}).

We examine the BRs of the decays of the $D$ assuming that $Z_2^H$ is mainly broken by the operators involving quarks and leptons of the third generation. In other words, we ignore all the Yukawa couplings $g_{ij1}$ and $\overline{g}_{ij1}$, except $g_{331}$ and $\overline{g}_{331}$,
so that the $D$ decays only into $t\tau$ and $b\nu$, and as a result $\mbox{{\rm BR}}(D \to b\nu)=1-{\rm BR}(D \to t\tau)$. To compute these BRs we incorporated the model in the Mathematica package SARAH v4.14.4~\cite{Staub:2008uz,Staub:2013tta,Goodsell:2014bna,Goodsell:2015ira,Staub:2015kfa,Goodsell:2016udb,Braathen:2017izn}, which evaluates the interaction vertices, and writes down model files for the FORTRAN code SPheno v4.0.5~\cite{Porod:2003um,Porod:2011nf} for performing phenomenological studies. 

\begin{figure}[tbp]
  \begin{center}
    \resizebox{12cm}{!}{\includegraphics{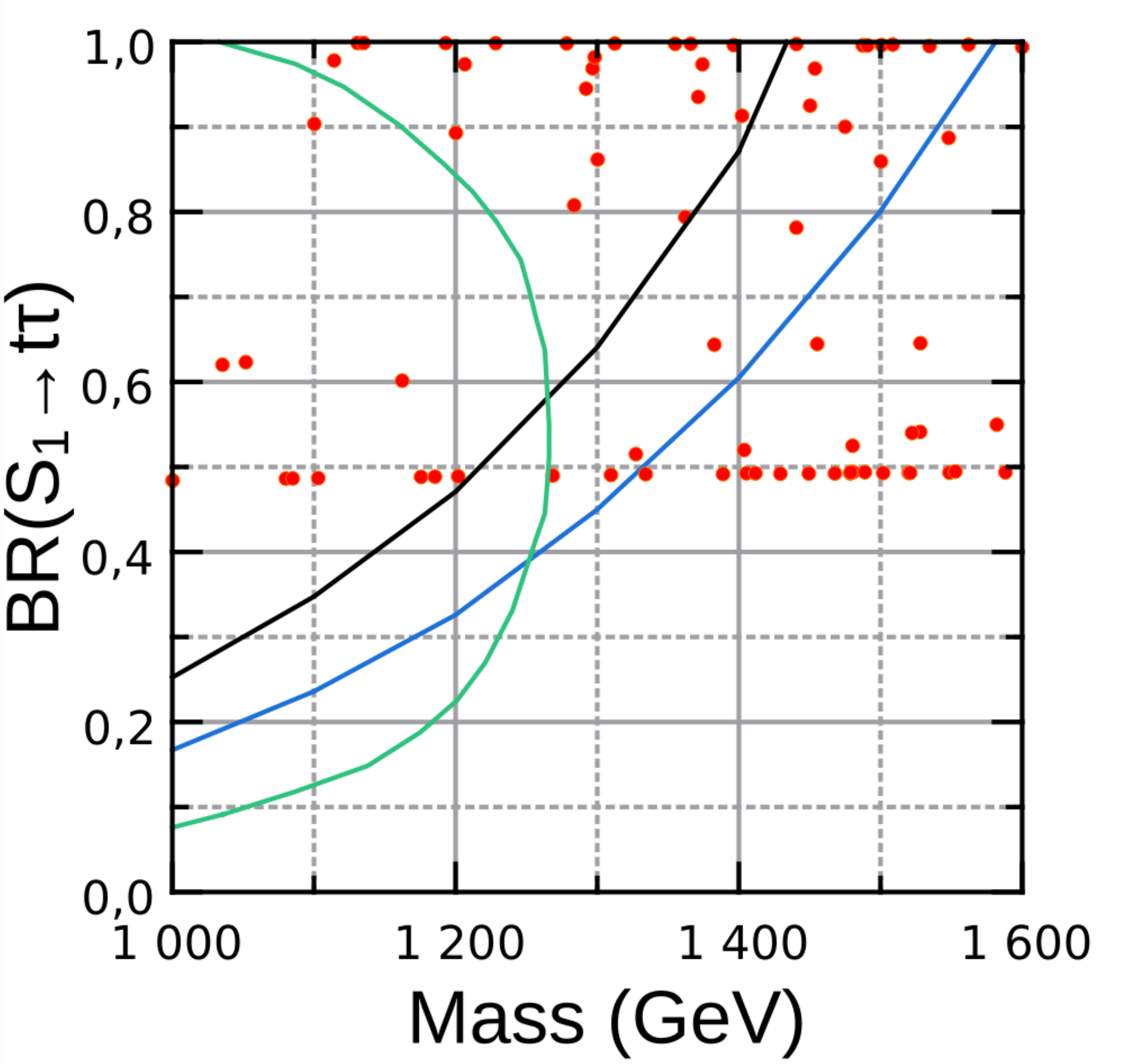}}
     \caption{The exclusion bound (blue) in the \{$m_{S_1}$,\,${\rm BR}(D \to t\tau)$\} plane. The area on the left hand side of the curves is excluded. The black and green lines are from the ATLAS analyses \cite{ATLAS:2021oiz} and \cite{ATLAS:2021jyv}, respectively. The red points correspond to some particular configurations of the E$_6$SSM parameter space, explained in the text.}
    \label{fig:excl}
  \end{center}
\end{figure}

In Fig. \ref{fig:excl}, we show the expected exclusion contour (blue) from our signal-to-background analysis in the \{$m_{S_1}$,\,${\rm BR}(D \to t\tau)$\} plane, together with those from two recent ATLAS results (black and green). Also plotted are a number of E$_6$SSM parameter space points, obtained for different values of $m^2_{D_1}$, $m^2_{\overline{D}_1}$, and $\kappa_1$. When the $D$ is predominantly the scalar component of the $\overline{D}_1$, it couples to the doublets and thus has ${\rm BR}(D \to t\tau)$ close to $50\%$. When it is mostly the $D_1$ it couples to singlets and decays solely to $t\tau$ if the right-handed neutrino is heavier than the $D$. Hence a lot of the points have the BR($D\to t\tau$) close to either $50\%$ or $100\%$. In the intermediate cases $M_{11}^2(1)\approx M_{22}^2(1)$, and there can be sizeable mixing between the scalar components of the $D_1$ and $\overline{D}_1$. 

We see in the figure that, when the $D$ is $\overline{D}_1$--like, our proposed analysis of the fully hadronic final state resulting from its pair-production can exclude its mass up to about 1340\,GeV. The exclusion bound is even stronger, close to 1580\,GeV, if the $D$ is predominantly the $D_1$ instead.


\section{\label{sec:concl} Summary and Conclusions}

LQs, being coloured, can be produced efficiently at the LHC, and are thus among the potential signatures of BSM physics. We have studied the possibility of searching third-generation LQs decaying to $t\tau$ in the fully hadronic channel. This analysis would complement the existing searches relying on the semi-leptonic channel, and has the advantage of larger statistics, as well as the possibility of reconstructing the mass of the LQ.

In the fully hadronic channel, we expect larger backgrounds from QCD processes. But we have shown that, with suitable event selection, the noise from the $t\bar{t}jj$ background gets under control, while the other backgrounds are negligible. We have found that this channel has a large estimated sensitivity for the Run 2 data, leading to an expected exclusion limit of up to $1580$~GeV (assuming BR$(S_1\rightarrow t\tau)=1$) on the LQ mass, thereby besting the ATLAS result based on the semi-leptonic channel by about $150$~GeV.

As an example of a model with such a signature, we have discussed a variant of the E$_{6}$SSM in which the fundamental representation contains scalar (and fermion) LQs. The $D$-term contributions to the scalar LQ masses are negative, unlike those to other sfermions, and they can hence be among the lightest BSM states of this SUSY scenario. If their couplings to the fermions of the first two generations are vanishing, one of the two lightest scalar LQs can decay dominantly to $t\tau$, and the other to $t\tau$ and $b\nu$, with nearly equal BR in each of these channels. We have then demonstrated how the fully hadronic channel is expected to improve the sensitivity of the LHC searches to the pair-production of the lightest LQ in this model.


\section*{Acknowledgments}
\noindent
The work of SK is partially supported by STDF under grant number 37272. SMo is supported in part through the NExT Institute and the STFC Consolidated Grant ST/L000296/1. HW is supported by the Carl Trygger Foundation under grant No. CTS18:164. MA acknowledges the receipt of the grant from the Abdus Salam International Centre for Theoretical Physics, Trieste, Italy.


\providecommand{\href}[2]{#2}\begingroup\raggedright\endgroup

\end{document}